\documentclass[12pt]{iopart}

\usepackage[latin9]{inputenc}

\usepackage{iopams}  
\usepackage{setstack}
\usepackage{color}

\usepackage{graphicx}

\begin{document}

\title[Analytical solution for multisingular vortex Gaussian beams]{Analytical solution for multisingular vortex Gaussian beams: The
mathematical theory of scattering modes}

\author{A. Ferrando}
\address{Departament d'Òptica. Interdisciplinary Modeling Group, \emph{InterTech.
}Universitat de València, Burjassot (València), Spain}

\author{M. A. García-March}
\address{ICFO-Institut de Ciències Fotòniques, The Barcelona Institute of
Science and Technology, Av. C.F. Gauss 3, 08860 Castelldefels (Barcelona),
Spain}

\begin{abstract}
We present a novel procedure to solve the Schrödinger equation, which
in optics is the paraxial wave equation, with an initial multisingular
vortex Gaussian beam. This initial condition has a number of singularities
in a plane transversal to propagation embedded in a Gaussian beam.
We use the scattering modes, which are solutions of the paraxial wave
equation that can be combined straightforwardly to express the initial
condition and therefore permit to solve the problem. To construct
the scattering modes one needs to obtain a particular set of polynomials, which
play an analogous role than Laguerre polynomials for Laguerre-Gaussian
modes. We demonstrate here the recurrence relations needed to determine
these polynomials. To stress the utility and strength of the method
we solve first the problem of an initial Gaussian beam with two positive
singularities and a negative one embedded in. We show that the solution
permits one to obtain analytical expressions. These can used to obtain
closed expressions for meaningful quantities, like the distance at
which the positive and negative singularities merge, closing the loop
of a vortex line. Furthermore, we present an example of calculation
of an specific discrete-Gauss state, which is the solution of the
diffraction of a Laguerre-Gauss state showing definite angular momentum
(that is, a highly charged vortex) by a thin diffractive element showing
certain discrete symmetry. We show that thereby this problem is solved
in a much simpler way than using the previous procedure based in the
integral Fresnel diffraction method. 
\end{abstract}
\maketitle

\section{Introduction}

The paraxial scalar wave equation for an optical field, which is formally
a two-dimensional (2D) linear Schrödinger equation, admits solutions
with quantized orbital angular momentum (OAM) \cite{1992:Allen,2000:Padgett,2007:Molina-Terriza,2008:Franke-Arnold,2011:Yao}.
These solutions present an on-axis phase singularity in the 2D transversal
plane, which is a zero in the intensity of the complex scalar wave,
with an undetermined phase. When considering a small circuit around
a phase singularity, the phase increases in an integer multiple of
$2\pi$. This integer is known as topological charge. The Laguerre-Gauss
(LG) modes are the mathematical solutions of such a paraxial equation
in free space showing OAM and, therefore, an on-axis phase singularity.
They are eigenstates of the OAM operator and, consequently, also
of the $O(2)$ continuous rotation group operator \cite{1986:Siegman},
and are usually termed as optical vortices. 

But the paraxial wave equation admits solutions with more intricate
phase profiles, like solutions showing many singularities. The whole
trajectory of a phase singularity along the evolution variable is
a vortex line, which can have very different geometries when considered
as independent entities, showing that the singularities present nontrivial
dynamics. Many such solutions, built by combining LG modes have been
reported \cite{1993:Indebetouw,2001:Jenkins,2001:Chavez,2004:Bandres,2004:Bandresb,2006:Volyar,2008:Deng,2008:Gutierrez,2012:Fadeyeva,2012:Steuernagel,2013:Lopez-Mago,2013:Martinez}.
For example, solutions with an intricate vortex line structure, forming
knots and loops can be obtained by superposition of LG modes \cite{2001:Berry,2004:Leach,2005:Leach,2010:Dennis}.
An  specific  formalism for the propagation of multisingular vortex beams, also named   
polynomial Gaussian beams, was presented in Refs.~\cite{2006:Roux,2008:Roux}.

Here, we present an easy, general and systematic procedure to construct a solution of the
paraxial wave equation with an initial condition showing any multisingular
structure, that is, a multisingular vortex beam. The key point is
to express the initial condition, showing a combination of singularities
embedded in a Gaussian beam, in terms of the so-called scattering
modes, introduced in Ref. \cite{2014:Ferrando}. The scattering modes
are solutions of the paraxial wave equation in free space that permit to
solve this equation straightforwardly.

We offer two examples to illustrate
the procedure: the first one exemplifies the procedure for a general
initial condition with a symmetric or non-symmetric structure of singularities.
To this end, we detail the particular case of an initial condition
with two single, positively charged off-axis singularities and a negative
charge located in the axis. We show that the expressions that solve
the problem along the whole evolution permit to obtain analytical
expressions for some meaningful quantities associated to the intricate
form of the corresponding vortex lines. For example, we show that it 
is possible to obtain an expression for the value of the distance
at which the singularities merge, that is, the distance at which the
vortex line loop closes. This can be obtained when initially the two
off-axis singularities are at the same distance of the origin. This
procedure represents a new strategy to investigate vortex knots and
loops in singular optics with analytical tools. In the second example
we consider an initial condition showing a definite discrete rotational
symmetry. This can be obtained as a diffraction of an LG mode showing
definite angular momentum (that is, a highly charged vortex) by a
thin diffractive discrete element showing certain discrete symmetry,
as described in \cite{2014:Ferrando,2013:Ferrando}. The particular
generation of these solutions ---termed as discrete Gaussian beams---
out of LG modes can show the phenomenon known as vortex transmutation,
that is the inversion of the topological charge as a consequence of
the discrete symmetry \cite{2005:Ferrando,2007:Garcia-Marchb,2009:Garcia-March,2009:Zacares,2012:Gao,2014:Novoa}. 

The procedure introduced in this paper 
is of interest in many different problems where the phase singularities play an important role. In particular, in the study of optical waves showing phase singularities
which is a large field in optics, called singular optics or, in the presence of nonlinearity, nonlinear singular optics \cite{1998:Soskin,2001:Soskin}. 
In three dimensions, a random light field is known as a laser speckle, where the vortex lines form complicated tangles and/or knots~\cite{2000BerryPRSA}. This can be treated mathematically with superpositions of planes waves~\cite{1985Goodman,2006HolleranOE}. Alternatively, they can also be analyzed by means of a basis of Laguerre-Gauss modes~\cite{1992DAlessandroOC}. In this sense the procedure introduced here should be considered an alternative to these approaches. 
One of the interesting fields in which we expect the  procedure introduced here would be fruitful is the study of optical knots in light beams~\cite{2001BerryPRSLA,2001BerryJPA,2011PadgettCP,2004LennisNature,2010DennisNatPhys}.
Also, it is of applicability in superfluids, particularly  in Bose-Einstein condensates (BEC), as the system is described by the Gross-Pitaevskii equation which  is formally identical to the 
 nonlinear Schr\"odinger equation. 
The recent realization of quantum knots in a BEC~\cite{2016HallNatPhys} makes this direction a very interesting venue to explore. In many other systems the concept of knot is of crucial importance and we believe that our method can be of great interest (see, e.g., examples in nematic colloids or water~\cite{2011TkalecScience,2013KlecknerNatPhys}). 
In addition, the direct experimental creation of electron-vortex beams carrying OAM proven in recent years shows the feasibility of using our analytical tools also in the context of electron quantum mechanics \cite{2013ClarkPRL}.

The paper is organized as follows. In section \ref{sec:Scattering-modes}
we introduce the scattering modes and describe the general procedure
to use them to solve the paraxial wave equation. In Sec. \ref{sec:Recurrence-relations-for-F-pol}
we discuss the $F$-polynomials, which are necessary to construct
the scattering modes, playing indeed an analogous role as Laguerre
polynomials for LG states. In Sec. \ref{sec:Demonstration-of-recurrence}
we demonstrate the recurrence relations necessary to construct these
polynomials, and therefore, any scattering mode. Sections \ref{sec:Example-1}
and \ref{sec:Example-2} deal with the two aforementioned examples.
We end this article offering our conclusions in Sec. \ref{sec:Conclusions}.

\section{Scattering modes \label{sec:Scattering-modes}}

Scattering modes are solutions of the paraxial diffraction equation 

\begin{equation}
-2ik_{0} \frac{\partial\phi}{\partial z}+\nabla_{t}^{2}\phi=0,\label{eq:par_diff_eq}
\end{equation}
where $\nabla_{t}\equiv(\partial/\partial x,\partial/\partial y)$
is the transverse gradient operator and $k_{0}$ is the light wavenumber,
verifying the following initial condition at $z=0$:
\begin{equation}
\Phi_{lp}\left(r,\theta,0\right)=r^{|l|+2p}\exp\left(il\theta\right)\exp\left[-\frac{k_{0}r^{2}}{2z_{R}}\right]\label{eq:init_cond_SM}
\end{equation}
where $z_{R}$ is the Rayleigh length. The radial exponential in (\ref{eq:init_cond_SM})
is nothing but $\phi_{00}(x,y,0)$, the fundamental Laguerre-Gauss
mode evaluated at $z=0$. In the complex variables $w=x+iy$ and $\overline{w}=x-iy$
this initial condition has the form:
\begin{eqnarray}
\Phi_{lp}\left(w,\overline{w},0\right) & = & w^{|l|}|w|^{2p}\exp\left[-\frac{k_{0}|w|^{2}}{2z_{R}}\right]\,\,\,\,\, l\ge0\nonumber\\
\Phi_{lp}\left(w,\overline{w},0\right) & = & \overline{w}^{|l|}|w|^{2p}\exp\left[-\frac{k_{0}|w|^{2}}{2z_{R}}\right]\,\,\,\,\, l<0.\label{eq:init_cond_SM2}
\end{eqnarray}

Let us briefly discuss how to construct solutions of Eq.~(\ref{eq:par_diff_eq}) with initial condition~(\ref{eq:init_cond_SM}). Let   $\hat w$ and $\hat{\overline{w}}$ be the complex position operators with associated momenta $\hat p=-i\partial/\partial w$ and $\hat{\overline{p}}=-i\partial/\partial \overline{w}$. These operators obey standard commutation relations $[\hat w,\hat p]=[\hat{\overline{w}},\hat{ \overline{p}}]=i$. They allow to write the Hamiltonian associated to Eq.~(\ref{eq:par_diff_eq}) as $\hat H=\hat p \hat{\overline{p}}$. Then, the evolution operator,  $\hat U(z)=\exp[i (2z/k_0) \hat H]$, fulfills 
\begin{equation}
[\hat w,\hat U(z)]=-(2z/k_0)\hat{\overline{p}}\hat U(z)\,\,\,\mbox{and}\,\,\, [\hat{\overline{w}},\hat U(z)]=-(2z/k_0)\hat{p}\hat U(z)\label{eq:commutator}
\end{equation}
Let us define the operators
\begin{equation}
\hat{l}_{+}(z)\equiv w-i\frac{2z}{k_0}\frac{\partial}{\partial\overline{w}}
\end{equation}
 and 
\begin{equation}
\hat{l}_{-}(z)\equiv\overline{w}-i\frac{2z}{k_0}\frac{\partial}{\partial w}.\label{eq:def_raise_low_op-1}
\end{equation}
Hence, by using the commutation relations~(\ref{eq:commutator}), the solution of Eq.~(\ref{eq:par_diff_eq}) is obtained by replacing $w$ and $\overline{w}$  in Eq.~(\ref{eq:init_cond_SM2}) by $\hat{l}_{+}$ and $\hat{l}_{-}$, respectively. One obtains 
\begin{eqnarray}
\Phi_{lp}\left(w,\overline{w},z\right) & = &\hat{l}_{+}^{|l|+p} \hat{l}_{-}^{p}\bar{\phi}_{00}(w,\overline{w},z)\,\,\,\,\, l\ge0\nonumber\\
\Phi_{lp}\left(w,\overline{w},z\right) & = & \hat{l}_{-}^{|l|+p} \hat{l}_{+}^{p}\bar{\phi}_{00}(w,\overline{w},z)\,\,\,\,\, l<0.\label{eq:sol1}
\end{eqnarray}
where the explicit expression for $\bar{\phi}_{00}$ in complex coordinates 
is 
\[
\bar{\phi}_{00}(|w|^{2},z)=\left(\frac{iz_{R}}{q(z)}\right)\exp\left(-\frac{i\pi w\overline{w}}{q(z)}\right).
\]
Note that $\bar{\phi}_{00}$ is the fundamental Laguerre-Gauss mode which solves Eq.~(\ref{eq:par_diff_eq})~\cite{1986:Siegman}. The solution~(\ref{eq:sol1}) can be written as 
\begin{eqnarray}
\Phi_{lp}(w,\overline{w},z) & = & \hat{l}_{\mathrm{sgn(l)}}^{|l|}(z)\,\left(\hat{l}_{+}^{p}\hat{l}_{-}^{p}\right)(z)\,\bar{\phi}_{00}(w,\overline{w},z)\nonumber \\
 & = & \hat{l}_{\mathrm{sgn(l)}}^{|l|}\,\hat{\triangle}^{p}(z)\,\bar{\phi}_{00}(w,\overline{w},z),\label{eq:SM_lp}
\end{eqnarray}
where we have introduced the ``diagonal'' operator $\hat{\Delta}\equiv\hat{l}_{+}\hat{l}_{-}$. We anticipate here, and justify below, that $\hat{l}_{+}$ and $\hat{l}_{+}$  are the raising and lowering operators for the angular momentum quantum number $l$, respectively,  while $\hat{\Delta}$ is the raising operator for the radial quantum number $p$. Repeteadly application of these operators to $\phi_{00}$ generates the solution with the corresponding values of $l$ and $p$. 

Alternatively, one can write Eqs.~(\ref{eq:sol1})  as 
\begin{eqnarray}
\Phi_{n\overline{n}}(w,\overline{w},z) & = & \hat{l}_{+}^{n}(z)\,\hat{l}_{-}^{\overline{n}}(z)\,\bar{\phi}_{00}(w,\overline{w},z).\nonumber \\
 & = & \left(w-i\frac{2 z}{k_0}\frac{\partial}{\partial\overline{w}}\right)^{n}\left(\overline{w}-i\frac{2 z}{k_0}\frac{\partial}{\partial w}\right)^{\overline{n}}\bar{\phi}_{00}(|w|^{2},z),\label{eq:def_SM}
\end{eqnarray}
where one identifies the angular momentum carried by the scattering mode $\phi_{n\overline{n}}$
is given by 
\begin{equation}
l=n-\overline{n}\label{eq:def_l}
\end{equation}
whereas 
\begin{equation}
p=\min\left(n,\overline{n}\right)\label{eq:def_p}. 
\end{equation}
The form of the solution~(\ref{eq:def_SM}) is that of the scattering
mode introduced in~\cite{2014:Ferrando}. The scattering
mode $\phi_{n\overline{n}}$ is obtained by applying $n$ times the $\hat{l}_{+}$
operator and $\overline{n}$ times the $\hat{l}_{-}$ one onto the fundamental
LG mode $\phi_{00}$.

In the following two sections we will justify that a general closed expression ---valid at \emph{any} $z$--- for the scattering
mode $\Phi_{lp}$ can be obtained. After introducing the Gaussian beam parameter $q(z)\equiv z+iz_{R}$, this expression is
\begin{eqnarray}
\hspace{-1.7cm}\Phi_{lp}\left(r,\theta,z\right) & =\left(\frac{iz_{R}}{q(z)}\right)^{|l|+1}\left(\frac{2zz_{R}}{k_{0}q(z)}\right)^{p}F_{p}^{|l|}\left(\gamma(z)r^{2}\right)r^{|l|}\exp\left(il\theta\right)\exp\left[-i\frac{k_{0}}{2}\frac{r^{2}}{q(z)}\right]\label{eq:SM_analytical}
\end{eqnarray}
where $\gamma(z)=\left(k_{0}/2\right)z_{R}\left[zq(z)\right]^{-1}$
and $F_{p}^{|l|}$ is a polynomial of $p$th order. The expression
for the scattering mode $\Phi_{lp}$ is therefore fully determined
by the polynomial of $p$th order $F_{p}^{|l|}(x)$. Explicit expressions
for these polynomials can be obtained using the recurrence relations
detailed in Section \ref{sec:Recurrence-relations-for-F-pol}. Scattering
modes also admit a representation in terms of the complex variables
$w$ and $\overline{w}$. For $l\ge0$,
\begin{equation}
\hspace{-1.7cm}\Phi_{lp}\left(w,\overline{w},z\right)=w^{|l|}\left(\frac{iz_{R}}{q(z)}\right)^{|l|+1}\left(\frac{2zz_{R}}{k_{0}q(z)}\right)^{p}F_{p}^{|l|}\left(\gamma(z)|w|^{2}\right)\exp\left[-i\frac{k_{0}}{2}\frac{|w|^{2}}{q(z)}\right],\label{eq:SM_complex_l_pos}
\end{equation}
 and for $l<0$
\begin{equation}
\hspace{-1.7cm}\Phi_{lp}\left(w,\overline{w},z\right)=\overline{w}^{|l|}\left(\frac{iz_{R}}{q(z)}\right)^{|l|+1}\left(\frac{2zz_{R}}{k_{0}q(z)}\right)^{p}F_{p}^{|l|}\left(\gamma(z)|w|^{2}\right)\exp\left[-i\frac{k_{0}}{2}\frac{|w|^{2}}{q(z)}\right].\label{eq:SM_complex_l_neg}
\end{equation}

Recurrence relations for $F$-polynomials are demonstrated from the
definition of scattering modes in terms of the differential operators
in the complex plane\footnote{From now on, we consider all distances normalized to $\lambda$, which
is equivalent to set $\lambda=1$ or, equivalently, $k_{0}=2\pi$
in all previous equations.}.

Scattering modes have a particular simple form at $z=0$ when using
the complex coordinates $w$ and $\overline{w}$:
\begin{equation}
\Phi_{n\overline{n}}(w,\overline{w},0)=w^{n}\overline{w}^{\overline{n}}\bar{\phi}_{00}(|w|^{2},0).\label{eq:SM_nnbar}
\end{equation}
This property provides a simple method to calculate the diffracted
field of any field whose expression at $z=0$ can be given as a product
of a series (finite or infinite) in powers of $w$ and $\overline{w}$
times a Gaussian. Let $\phi$ be such a field. Then
\begin{equation}
\phi\left(w,\overline{w},0\right)=t\left(w,\overline{w}\right)\bar{\phi}_{00}(w,\overline{w},0)=\sum_{n,\overline{n}}t_{n\overline{n}}\left[w^{n}\overline{w}^{\overline{n}}\bar{\phi}_{00}(w,\overline{w},0)\right].\label{eq:arbitrary_field-at-zero}
\end{equation}
According to Ref. \cite{2014:Ferrando}, the value of the field at
arbitrary $z$ is obtained by the simple substitution rule: $w\rightarrow\hat{l}_{+}(z)$
and $\overline{w}\rightarrow\hat{l}_{-}(z)$. After substitution in
(\ref{eq:arbitrary_field-at-zero}), we immediately recognize that
the term in brackets in this expression becomes nothing but the scattering mode
$\Phi_{n\overline{n}}$. Thus, the coefficients of the expansion of
the function $t$ at $z=0$ are \emph{also } the coefficients of the
diffracted field in terms of the scattering modes valid in the entire space:
\begin{equation}
\phi\left(w,\overline{w},z\right)=\sum_{n,\overline{n}}t_{n\overline{n}}\Phi_{n\overline{n}}(w,\overline{w},z).\label{eq:expansion_t_coef_SM}
\end{equation}
Or, alternatively, using $l$ and $p$ as ``quantum numbers''
\[
\phi\left(w,\overline{w},z\right)=\sum_{lp}t_{lp}\Phi_{lp}(w,\overline{w},z),
\]
in which $l$ and $p$ are calculated using Eqs.~(\ref{eq:def_l})
and (\ref{eq:def_p}).

\section{Recurrence relations for $F$-polynomials\label{sec:Recurrence-relations-for-F-pol} }

The $F$-polynomials in equations (\ref{eq:SM_analytical})-(\ref{eq:SM_complex_l_neg})
play an analogous role as Laguerre polynomials for LG states. The
initial condition (\ref{eq:init_cond_SM}) is, however, different
from that fulfilled for LG states in the $p\ne0$ case. Therefore,
the scattering mode $\Phi_{lp}$ with $p\ne0$ is necessarily different
from any LG mode and so is its expression (\ref{eq:SM_analytical}).
This includes the $F$-polynomials, which are not Laguerre polynomials.
However, $F$-polynomials can be similarly provided by specific recurrence
relations as Laguerre ones.

We distinguish between \emph{fundamental} $F$-polynomials, characterized
by $l=0$ and thus described exclusively by the $p$-index, and \emph{generalized}
$F$-polynomials, for which $l\ne0$. We denote a polynomial of the
former type as $F_{p}\left(x\right)\equiv F_{p}^{0}(x)$ whereas we
reserve the full notation $F_{p}^{|l|}(x)$ for the generalized form
of the polynomial.

\subsection{Recurrence relation for fundamental $F$-polynomials }

The fundamental $F$-polynomials satisfy the following recurrence
set of differential equations:
\begin{equation}
\hspace{-1.cm} F_{p+1}\left(x\right)=\left(1-x\right)F_{p}\left(x\right)-\left(1-2x\right)\frac{dF{}_{p}\left(x\right)}{dx}-x\frac{d^{2}F_{p}(x)}{dx^{2}},\,\,\,\,\, p=0,1,2,\dots\label{eq:set_fund_F_diff_rec}
\end{equation}
with 
\[
\hspace{-1.cm}F_{0}(x)=1,
\]
which permits to solve the recurrence set~(\ref{eq:set_fund_F_diff_rec})
univocally. 

An equivalent construction of the fundamental $F$-polynomials can
be obtained by writing the polynomials in an explicit manner: 
\[
F_{p}\left(x\right)=\sum_{j=0}^{p}c_{j}^{p}x^{j}.
\]
In this way, the set of differential  equations~(\ref{eq:set_fund_F_diff_rec})
turns into the following system of algebraic recurrence relations for the coefficients
$c_{j}^{p}$:
\begin{equation}
c_{j}^{p+1}=\left(1+2j\right)c_{j}^{p}-\left(1+j\right)^{2}c_{j+1}^{p}-c_{j-1}^{p}\label{eq:rec_rel_coeff_fund}
\end{equation}
together with the conditions 
\begin{eqnarray*}
c_{j}^{p} & = & 0,\,\,\, j<0\,\,\,\mathrm{or}\,\,\, j>p\\
c_{p}^{p} & = & \left(-1\right)^{p},\,\,\, p=0,1,2,\dots,
\end{eqnarray*}
which act as initialization conditions for the recurrence chain. 

We provide in Table~\ref{tab:fund_pol} the explicit expressions
of the lower order fundamental polynomials ---up to sixth order---
obtained using the previous recurrence relations.

\begin{table}
\begin{tabular}{|l|}
\hline 
$F_{p}\left(x\right)$\tabularnewline
\hline 
\hline 
$F_{0}\left(x\right)=1$\tabularnewline
\hline 
$F_{1}\left(x\right)=1-x$\tabularnewline
\hline 
$F_{2}\left(x\right)=2-4x+x^{2}$\tabularnewline
\hline 
$F_{3}\left(x\right)=6-18x+9x^{2}-x^{3}$\tabularnewline
\hline 
$F_{4}\left(x\right)=24-96x+72x^{2}-16x^{3}+x^{4}$\tabularnewline
\hline 
$F_{5}\left(x\right)=120-600x+600x^{2}-200x^{3}+25x^{4}-x^{5}$\tabularnewline
\hline 
$F_{6}\left(x\right)=720-4320x+5400x^{2}-2400x^{3}+450x^{4}-36x^{5}+x^{6}$\tabularnewline
\hline 
\end{tabular}

\caption{Lower order fundamental $F$-polynomials.\label{tab:fund_pol}}
\end{table}

\subsection{Recurrence relation for generalized $F$-polynomials }

The generalized $F$-polynomials $F_{p}^{|l|}$ fulfill the following
recurrence set of differential equations:
\[
F_{p}^{|l|+1}(x)=F_{p}^{|l|}(x)-\frac{dF_{p}^{|l|}(x)}{dx},\,\,\,\,\,|l|=0,1,2,\dots,
\]
which together with the initial condition
\[
F_{p}^{0}(x)=F_{p}(x),
\]
permit to obtain recursively the generalized $F$-polynomial of order
$p$ and angular momentum $|l|$, $F_{p}^{|l|}(x)$, from the corresponding
fundamental $F$-polynomial $F_{p}(x)$ previously determined by means
of the recurrence relations (\ref{eq:set_fund_F_diff_rec}). It is
simple to deduce that $F_{0}^{|l|}(x)=F_{0}(x)=1$.

An equivalent recurrence relation for the coefficients of the generalized
$F$-polynomial
\[
F_{p}^{|l|}(x)=\sum_{j=0}^{p}c_{j}^{|l|,p}x^{j}
\]
is
\begin{equation}
c_{j}^{|l|+1,p}=c_{j}^{|l|,p}-\left(j+1\right)c_{j+1}^{|l|,p}\label{eq:rec_rel_gen_coeff}
\end{equation}
with
\[
c_{j}^{0,p}=c_{j}^{p},
\]
$c_{j}^{p}$ being the $j$-th coefficient of the corresponding fundamental
polynomial $F_{p}(x)$ previously evaluated using the recurrence relations
(\ref{eq:rec_rel_coeff_fund}). The recurrence relations (\ref{eq:rec_rel_gen_coeff})
permits to obtain eventually all the coefficients $c_{j}^{|l|,p}$
out of the fundamental ones $c_{j}^{p}$ in a recursive way.

In Table \ref{tab:generalized_pol} we include the lower order generalized
polynomials $F_{p}^{|l|}$ for angular momenta $|l|=1,2,\,\mathrm{and}\,3$.

\begin{table}
\begin{tabular}{|l|}
\hline 
$F_{p}^{1}\left(x\right)$\tabularnewline
\hline 
\hline 
$F_{0}^{1}\left(x\right)=1$\tabularnewline
\hline 
$F_{1}^{1}\left(x\right)=2-x$\tabularnewline
\hline 
$F_{2}^{1}\left(x\right)=6-6x+x^{2}$\tabularnewline
\hline 
$F_{3}^{1}\left(x\right)=24-36x+12x^{2}-x^{3}$\tabularnewline
\hline 
$F_{4}^{1}\left(x\right)=120-240x+120x^{2}-20x^{3}+x^{4}$\tabularnewline
\hline 
$F_{5}^{1}\left(x\right)=720-1800x+1200x^{2}-300x^{3}+30x^{4}-x^{5}$\tabularnewline
\hline 
$F_{6}^{1}\left(x\right)=5040-15120x+12600x^{2}-4200x^{3}+630x^{4}-42x^{5}+x^{6}$\tabularnewline
\hline 
\end{tabular}

\bigskip{}
\bigskip{}

\begin{tabular}{|l|}
\hline 
$F_{p}^{2}\left(x\right)$\tabularnewline
\hline 
\hline 
$F_{0}^{2}\left(x\right)=1$\tabularnewline
\hline 
$F_{1}^{2}\left(x\right)=3-x$\tabularnewline
\hline 
$F_{2}^{2}\left(x\right)=12-8x+x^{2}$\tabularnewline
\hline 
$F_{3}^{2}\left(x\right)=60-60x+15x^{2}-x^{3}$\tabularnewline
\hline 
$F_{4}^{2}\left(x\right)=360-480x+180x^{2}-24x^{3}+x^{4}$\tabularnewline
\hline 
$F_{5}^{2}\left(x\right)=2520-4200x+2100x^{2}-420x^{3}+35x^{4}-x^{5}$\tabularnewline
\hline 
$F_{6}^{2}\left(x\right)=20160-40320x+25200x^{2}-6720x^{3}+840x^{4}-48x^{5}+x^{6}$\tabularnewline
\hline 
\end{tabular}

\bigskip{}
\bigskip{}

\begin{tabular}{|l|}
\hline 
$F_{p}^{3}\left(x\right)$\tabularnewline
\hline 
\hline 
$F_{0}^{3}\left(x\right)=1$\tabularnewline
\hline 
$F_{1}^{3}\left(x\right)=4-x$\tabularnewline
\hline 
$F_{2}^{3}\left(x\right)=20-10x+x^{2}$\tabularnewline
\hline 
$F_{3}^{3}\left(x\right)=120-90x+18x^{2}-x^{3}$\tabularnewline
\hline 
$F_{4}^{3}\left(x\right)=840-840x+252x^{2}-28x^{3}+x^{4}$\tabularnewline
\hline 
$F_{5}^{3}\left(x\right)=6720-8400x+3360x^{2}-560x^{3}+40x^{4}-x^{5}$\tabularnewline
\hline 
$F_{6}^{3}\left(x\right)=60480-90720x+45360x^{2}-10080x^{3}+1080x^{4}-54x^{5}+x^{6}$\tabularnewline
\hline 
\end{tabular}

\caption{Lower order generalized $F$-polynomials $F_{p}^{|l|}(x)$ for the
lowest angular momenta: $|l|=1,2,3$. \label{tab:generalized_pol}}
\end{table}

\section{Demonstration of recurrence relations\label{sec:Demonstration-of-recurrence}}

\subsection{Raising and lowering operators}

The operators $\hat{l}_{+}$ and $\hat{l}_{-}$ as well as the diagonal
operator $\hat{\Delta}$ can be interpreted as lowering and raising
operators for the quantum numbers $l$ and $p$. The interpretation
is clearer when working in the $z=0$ plane since the expressions
for $\hat{l}_{+}$ and $\hat{l}_{-}$ in definition (\ref{eq:def_SM})
take the simple multiplicative form:
\begin{equation}
\Phi_{n\overline{n}}\left(w,\overline{w},0\right)=w^{n}\,\overline{w}^{\overline{n}}\,\bar{\phi}_{00}(w,\overline{w},0),\label{eq:SM_n_nbar}
\end{equation}
so that
\begin{eqnarray*}
\hat{l}_{+}(0)\Phi_{n\overline{n}}\left(w,\overline{w},0\right) & = & w^{n+1}\,\overline{w}^{\overline{n}}\,\bar{\phi}_{00}(w,\overline{w},,0)\\
 & = & \Phi_{n+1,\overline{n}}\left(w,\overline{w},0\right).
\end{eqnarray*}
Analogously,
\[
\hat{l}_{-}(0)\Phi_{n\overline{n}}\left(w,\overline{w},0\right)=\Phi_{n,\overline{n}+1}\left(w,\overline{w},,0\right).
\]
Therefore, $\hat{l}_{+}$ and $\hat{l}_{-}$ raise the value of the
index $n$ and $\overline{n}$ by one unit, respectively. If we use $ket$
notation to represent the scattering mode $\Phi_{n\overline{n}}$
as $\left|SM(n,\overline{n})\right\rangle $ this means: 
\begin{eqnarray*}
\hat{l}_{+}\left|SM(n,\overline{n})\right\rangle  & = & \left|SM(n+1,\overline{n})\right\rangle \\
\hat{l}_{-}\left|SM(n,\overline{n})\right\rangle  & = & \left|SM(n,\overline{n}+1)\right\rangle .
\end{eqnarray*}
In terms of the angular momentum $l=n-\overline{n}$, it is evident that
$\hat{l}_{+}$ increases its value by one unit whereas $\hat{l}_{-}$
decrease it in the same amount. So that, $\hat{l}_{+}$ and $\hat{l}_{-}$
are the angular momentum raising and lowering operators for scattering
modes. It is not difficult to check that the action of $\hat{l}_{+}$
and $\hat{l}_{-}$ onto a generic scattering mode written in terms
of $l$ and $p$ is given by:
\begin{eqnarray}
\hspace{-1.7cm}\hat{l}_{+}\left|SM(l,p)\right\rangle  & = & \left|SM(l+1,p)\right\rangle \,\,\,\,\,\,\,\,\hat{l}_{-}\left|SM(l,p)\right\rangle =\left|SM(l-1,p+1)\right\rangle \,\,\,\,\,\, l>0,\nonumber \\
\hspace{-1.7cm}\hat{l}_{+}\left|SM(l,p)\right\rangle  & = & \left|SM(l+1,p+1)\right\rangle \,\,\,\,\,\,\,\,\hat{l}_{-}\left|SM(l,p)\right\rangle =\left|SM(l-1,p)\right\rangle \,\,\,\,\,\, l\le0.\label{eq:lp_lm_raising_low_op}
\end{eqnarray}
Note that depending on the sign of $l$, the action of the angular
momentum raising and lowering operators can also affect the $p$ index
. However, the particular combination of these operators given by
the diagonal operator $\hat{\Delta}\equiv\hat{l}_{+}\hat{l}_{-}$
acts systematically as a raising operator for the $p$ index since
an increase in one unit for $n$ and $\overline{n}$ implies 
\[
p'=\min(n+1,\overline{n}+1)=\min(n,\overline{n})+1=p+1.
\]
Therefore
\begin{equation}
\hat{\Delta}\left|SM(l,p)\right\rangle =\left|SM(l,p+1)\right\rangle .\label{eq:delta_raising_op}
\end{equation}
Expressions~(\ref{eq:lp_lm_raising_low_op}) and (\ref{eq:delta_raising_op})
are the key elements to determine the recurrence relations for fundamental
and generalized $F$-polynomials.

\subsection{Derivation of the recurrence relation for fundamental $F$-polynomials}

Our starting point is Eq.~(\ref{eq:delta_raising_op}) for the diagonal
operator $\hat{\Delta}$ applied to a scattering mode with $l=0$.
In terms of the functions $\Phi_{0p}$ and $\Phi_{0p+1}$ this equation
reads
\begin{eqnarray}
\Phi_{0p+1}\left(w,\overline{w},z\right) & = & \hat{l}_{+}(z)\hat{l}_{-}(z)\Phi_{0p}\left(w,\overline{w},z\right)\nonumber \\
 & = & \left(w-i\frac{z}{\pi}\frac{\partial}{\partial\overline{w}}\right)\left(\overline{w}-i\frac{z}{\pi}\frac{\partial}{\partial w}\right)\Phi_{0p}\left(w,\overline{w},z\right).\label{eq:delta_on_SM_p}
\end{eqnarray}
Let us consider this equation for the lower order scattering modes. 

An explicit calculation of the derivatives for $p=0$ provides the
following result (taking into account that $\Phi_{00}=\bar{\phi}_{00}$
):
\begin{eqnarray*}
\Phi_{01}\left(w,\overline{w},z\right) & = & \left(\frac{iz_{R}}{q(z)}\right)\left(w-i\frac{z}{\pi}\frac{\partial}{\partial\overline{w}}\right)\left(\overline{w}-i\frac{z}{\pi}\frac{\partial}{\partial w}\right)\exp\left(-\frac{i\pi w\overline{w}}{q(z)}\right)\\
 & = & \left(\frac{zz_{R}}{\pi q(z)}\right)\left(1-x\right)\bar{\phi}_{00}(|w|^{2},z),
\end{eqnarray*}
where $x\equiv\pi z_{R}\left[zq(z)\right]^{-1}|w|^{2}$. 

For $p=1$, 
\begin{eqnarray*}
\Phi_{02}\left(w,\overline{w},z\right) & = & \left(w-i\frac{z}{\pi}\frac{\partial}{\partial\overline{w}}\right)\left(\overline{w}-i\frac{z}{\pi}\frac{\partial}{\partial w}\right)\Phi_{01}\left(w,\overline{w},z\right)\\
 & = & \left(\frac{zz_{R}}{\pi q(z)}\right)^{2}\left(2-4x+x^{2}\right)\bar{\phi}_{00}(|w|^{2},z).
\end{eqnarray*}
Thus, for arbitrary $p$ we expect the following structure: 
\begin{equation}
\Phi_{0p}\left(w,\overline{w},z\right)=\left(\frac{zz_{R}}{\pi q(z)}\right)^{p}F_{p}\left[\gamma(z)|w|^{2}\right]\bar{\phi}_{00}(|w|^{2},z),\label{eq:fund_SM_expr_F(x)}
\end{equation}
where $F_{p}(x)$ is a polynomial of order $p$ in $x=\gamma(z)|w|^{2}$,
in which we have defined $\gamma(z)=\pi z_{R}\left[zq(z)\right]^{-1}$. 

After substituting the \emph{ansatz }(\ref{eq:fund_SM_expr_F(x)})
in Eq.~(\ref{eq:delta_on_SM_p}), we obtain an explicit expression
for the fundamental $F$-polynomial of order $p+1$ in terms of $F_{p}$
and its derivatives:
\begin{eqnarray*}
\hspace{-1.8cm}F_{p+1}\left[\gamma(z)|w|^{2}\right] & = & \left(\frac{zz_{R}}{\pi q(z)}\right)^{-1}\left[\bar{\phi}_{00}(|w|^{2},z)\right]^{-1}\times\\
\hspace{-1.8cm} &  & \left(w-i\frac{z}{\pi}\frac{\partial}{\partial w}\right)\left(\overline{w}-i\frac{z}{\pi}\frac{\partial}{\partial\overline{w}}\right)\left\{ F_{p}\left[\gamma(z)|w|^{2}\right]\bar{\phi}_{00}(|w|^{2},z)\right\} .
\end{eqnarray*}
Developing the differential operators in terms of the $\partial/\partial w$
and $\partial/\partial\overline{w}$ derivatives and acting on the product
$F_{p}\bar{\phi}_{00}$ provide the following result:
\begin{eqnarray*}
\hspace{-1.8cm}F_{p+1}\left[\gamma(z)|w|^{2}\right]&=&\left[zq(z)\right]^{-1}\left\{\left(zq(z)-\pi z_{R}w\overline{w}\right)F_{p}\left[\gamma(z)|w|^{2}\right]\right.\\
\hspace{-1.8cm} & -&\left.\left(zq(z)-2\pi z_{R}w\overline{w}\right)F_{p}'\left[\gamma(z)|w|^{2}\right]-\left(\pi z_{R}w\overline{w}\right)F_{p}''\left[\gamma(z)|w|^{2}\right]\right\}.
\end{eqnarray*}
After reintroducing the argument of the $F$-polynomial as $x=\gamma(z)w\overline{w}=\pi z_{R}\left[zq(z)\right]^{-1}w\overline{w}$,
we obtain the desired recurrence relation in terms of the single variable
$x$:
\[
F_{p+1}\left(x\right)=\left(1-x\right)F_{p}\left(x\right)-\left(1-2x\right)\frac{dF_{p}\left(x\right)}{dx}-x\frac{d^{2}F_{p}\left(x\right)}{dx^{2}}.
\]
Since according to Eq.~(\ref{eq:fund_SM_expr_F(x)}) $\Phi_{00}=\bar{\phi}_{00}$,
we additionally have that $F_{0}(x)=1$, which is the initial condition
for the recurrence chain. Note how the $F-$polynomials of the lower
order scattering modes we used to set the \emph{ansatz }(\ref{eq:fund_SM_expr_F(x)})\emph{
} are recovered using the previous recurrence relation, thus showing
the consistency of the procedure.

\subsection{Derivation of the recurrence relation for generalized $F$-polynomials}

We start by proving the relation for $l>0$. We consider the first
relation in Eq.~(\ref{eq:lp_lm_raising_low_op}) relating the scattering mode 
$\left|SM(l+1,p)\right\rangle $ with $\left|SM(l,p)\right\rangle $
by means of the angular momentum raising operator $\hat{l}_{+}$.
In terms of the functions $\Phi_{l+1,p}$ and $\Phi_{lp}$ this equation
reads:
\begin{eqnarray}
\Phi_{l+1,p}\left(w,\overline{w},z\right) & = & \hat{l}_{+}(z)\Phi_{lp}\left(w,\overline{w},z\right)\nonumber \\
 & = & \left(w-i\frac{z}{\pi}\frac{\partial}{\partial\overline{w}}\right)\Phi_{lp}\left(w,\overline{w},z\right).\label{eq:lp_on_SM_lp}
\end{eqnarray}

On the other hand, the general form of $\Phi_{lp}$ in terms of $w$
and $\overline{w}$ is given by symmetry considerations and by the
action of derivatives, as explained in Ref. \cite{2014:Ferrando}.
In this reference, it was shown that a scattering mode with $l>0$ has well-defined
angular momentum $l$ and, thus, it transforms as $\Phi_{lp}\rightarrow\left(\exp il\theta\right)\Phi_{lp}$
under the continuous rotation $w\rightarrow\left(\exp i\theta\right)w$.
So that, $\Phi_{lp}$ has to be proportional to $w^{l}$ times a function
dependent exclusively on the $O(2)$ invariant $|w|^{2}$. Since both
$\partial/\partial w$ and $\partial/\partial\overline{w}$ derivatives
of any order acting on the fundamental Gaussian mode $\bar{\phi}_{00}$
yield terms proportional to $\bar{\phi}_{00}$,
this $O(2)$ invariant function has to be also proportional to $\bar{\phi}_{00}$.
Thus, the general form for the scattering mode in terms of $w$ and $\overline{w}$
is given by $\Phi_{lp}\sim w^{l}F_{p}^{l}\bar{\phi}_{00}$,
where $F_{p}^{l}$ is, up to this point, a function (to be determined)
dependent on $|w|^{2}$. The normalization factor is, in general,
$z$-dependent. In the same way as we did for the fundamental $F$-polynomials,
a calculation of the lower order scattering modes using Eq.~(\ref{eq:lp_on_SM_lp})
helps us to find an \emph{ansatz} consistent with this symmetry argument.
The calculation provides the following results. For $p=1$ and $l=0$:
\begin{eqnarray*}
\hspace{-1.7cm}\Phi_{11}\left(w,\overline{w},z\right) & = & \hat{l}_{+}(z)\Phi_{01}\left(w,\overline{w},z\right)=\left(w-i\frac{z}{\pi}\frac{\partial}{\partial\overline{w}}\right)\Phi_{01}\left(w,\overline{w},z\right)\\
\hspace{-1.7cm} & = & w\left(\frac{iz_{R}}{q(z)}\right)\left(\frac{zz_{R}}{\pi q(z)}\right)\left(2-x\right)\bar{\phi}_{00}(|w|^{2},z).
\end{eqnarray*}
For $p=1$ and $l=1$:
\begin{eqnarray*}
\hspace{-1.7cm}\Phi_{21}\left(w,\overline{w},z\right) & = & \hat{l}_{+}(z)\Phi_{11}\left(w,\overline{w},z\right)=\left(w-i\frac{z}{\pi}\frac{\partial}{\partial\overline{w}}\right)\Phi_{11}\left(w,\overline{w},z\right)\\
\hspace{-1.7cm} & = & w^{2}\left(\frac{iz_{R}}{q(z)}\right)^{2}\left(\frac{zz_{R}}{\pi q(z)}\right)\left(3-x\right)\bar{\phi}_{00}(|w|^{2},z).
\end{eqnarray*}
For $p=1$ and $l=2$:
\begin{eqnarray*}
\hspace{-1.7cm}\Phi_{22}\left(w,\overline{w},z\right) & = & \hat{\Delta}(z)\Phi_{21}\left(w,\overline{w},z\right)=\left(w-i\frac{z}{\pi}\frac{\partial}{\partial\overline{w}}\right)\left(\overline{w}-i\frac{z}{\pi}\frac{\partial}{\partial w}\right)\Phi_{21}\left(w,\overline{w},z\right)\\
 \hspace{-1.7cm}& = & w^{2}\left(\frac{iz_{R}}{q(z)}\right)^{2}\left(\frac{zz_{R}}{\pi q(z)}\right)^{2}\left(12-8x+x^{2}\right)\bar{\phi}_{00}(|w|^{2},z).
\end{eqnarray*}
This calculation sets the following \emph{ansatz} for the scattering mode
with $l=|l|>0$ represented by $\Phi_{lp}$ :
\begin{equation}
\Phi_{|l|p}\left(w,\overline{w},z\right)=w^{|l|}\left(\frac{iz_{R}}{q(z)}\right)^{|l|}\left(\frac{zz_{R}}{\pi q(z)}\right)^{p}F_{p}^{|l|}\left(\gamma(z)|w|^{2}\right)\bar{\phi}_{00}(|w|^{2},z).\label{eq:SM_complex_w}
\end{equation}
Note that this equation provides a simple reduction to the $l=0$
case ---Eq.~(\ref{eq:fund_SM_expr_F(x)})--- by naturally assuming
that $F_{p}(x)=F_{p}^{0}(x)$.

Now, the equation for the angular momentum raising operator $\hat{l}_{+}$
(\ref{eq:lp_on_SM_lp}) permits to write an explicit expression for
the generalized $F$-polynomial $F_{p}^{|l|+1}$ in terms of $F_{p}^{|l|}$
and its derivative:
\begin{eqnarray*}
F_{p}^{|l|+1}\left[\gamma(z)w\overline{w}\right] & = & w^{-1}\left(\frac{iz_{R}}{q(z)}\right)^{-1}\left[\bar{\phi}_{00}(w,\overline{w},z)\right]^{-1}\times\\
 &  & \left(w-i\frac{z}{\pi}\frac{\partial}{\partial\overline{w}}\right)\left\{ F_{p}^{|l|}\left(\gamma(z)|w|^{2}\right)\bar{\phi}_{00}(|w|^{2},z)\right\} .
\end{eqnarray*}
Developing the derivative of the $F_{p}^{|l|}\bar{\phi}_{00}$
product and introducing $x=\gamma(z)w\overline{w}=\pi z_{R}\left[zq(z)\right]^{-1}|w|^{2}$
in the previous equation provides the following relation :
\begin{eqnarray}
F_{p}^{|l|+1}\left(x\right) & = & F_{p}^{|l|}\left(x\right)-\frac{zq\left(z\right)}{\pi z_{R}}\gamma\left(z\right)\frac{dF_{p}^{|l|}\left(x\right)}{dx}\nonumber \\
 & = & F_{p}^{|l|}\left(x\right)-\frac{dF_{p}^{|l|}\left(x\right)}{dx}.\label{eq:recur_rel_gen_F-polynom}
\end{eqnarray}

An analogous calculation applies to a scattering mode with $l=-|l|<0$. Now
we use, instead of Eq.~(\ref{eq:lp_on_SM_lp}), the second relation
for the lowering operator $\hat{l}_{-}$ in Eqs.~(\ref{eq:lp_lm_raising_low_op})
valid for $l<0$ 
\begin{eqnarray}
\hspace{-1.7cm}\Phi_{l-1,p}\left(w,\overline{w},z\right) & = & \hat{l}_{-}(z)\Phi_{lp}\left(w,\overline{w},z\right)\nonumber \\
\hspace{-1.7cm} & = & \left(\overline{w}-i\frac{z}{\pi}\frac{\partial}{\partial w}\right)\Phi_{lp}\left(w,\overline{w},z\right).\label{eq:lm_on_SM_lp}
\end{eqnarray}
 A straightforward calculation of the lower order scattering modes provides the
following results. For $p=1$ and $l=0$: 
\begin{eqnarray*}
\hspace{-1.7cm}\Phi_{-11}\left(w,\overline{w},z\right) & = & \hat{l}_{-}(z)\Phi_{01}\left(w,\overline{w},z\right)=\left(\overline{w}-i\frac{z}{\pi}\frac{\partial}{\partial w}\right)\Phi_{01}\left(w,\overline{w},z\right)\\
\hspace{-1.7cm} & = & \overline{w}\left(\frac{iz_{R}}{q(z)}\right)\left(\frac{zz_{R}}{\pi q(z)}\right)\left(2-x\right)\bar{\phi}_{00}(|w|^{2},z).
\end{eqnarray*}
For $p=1$ and $l=-1$ 
\begin{eqnarray*}
\hspace{-1.7cm}\Phi_{-21}\left(w,\overline{w},z\right) & = & \hat{l}_{-}(z)\Phi_{-11}\left(w,\overline{w},z\right)=\left(\overline{w}-i\frac{z}{\pi}\frac{\partial}{\partial w}\right)\Phi_{-11}\left(w,\overline{w},z\right)\\
\hspace{-1.7cm} & = & \overline{w}^{2}\left(\frac{iz_{R}}{q(z)}\right)^{2}\left(\frac{zz_{R}}{\pi q(z)}\right)\left(3-x\right)\bar{\phi}_{00}(|w|^{2},z).
\end{eqnarray*}
For $p=1$ and $l=-2$ 
\begin{eqnarray*}
\hspace{-1.7cm}\Phi_{-22}\left(w,\overline{w},z\right) & = & \hat{\Delta}(z)\Phi_{-21}\left(w,\overline{w},z\right)=\left(w-i\frac{z}{\pi}\frac{\partial}{\partial\overline{w}}\right)\left(\overline{w}-i\frac{z}{\pi}\frac{\partial}{\partial w}\right)\Phi_{-21}\left(w,\overline{w},z\right)\\
\hspace{-1.7cm} & = & \overline{w}^{2}\left(\frac{iz_{R}}{q(z)}\right)^{2}\left(\frac{zz_{R}}{\pi q(z)}\right)^{2}\left(12-8x+x^{2}\right)\bar{\phi}_{00}(|w|^{2},z).
\end{eqnarray*}
We immediately recognize that this structure is identical to that
of the $l>0$ scattering modes with the exception of the dependence on $\overline{w}^{|l|}$
instead of on $w^{|l|}$. Thus, the general form for $\Phi_{lp}$
for $l=-|l|<0$ is:
\begin{equation}
\Phi_{-|l|p}\left(w,\overline{w},z\right)=\overline{w}^{|l|}\left(\frac{iz_{R}}{q(z)}\right)^{|l|}\left(\frac{zz_{R}}{\pi q(z)}\right)^{p}F_{p}^{|l|}\left(\gamma(z)|w|^{2}\right)\bar{\phi}_{00}(|w|^{2},z).\label{eq:SM_complex_wc}
\end{equation}
Therefore $F_{p}^{-|l|}=F_{p}^{|l|}$ and the recurrence relation
for generalized $F$-polynomial is also given by Eq.~(\ref{eq:recur_rel_gen_F-polynom}).

Note, as before, that the polynomials of the lower order scattering modes obtained by
direct derivation applying the definitions (\ref{eq:lp_on_SM_lp})
and (\ref{eq:lm_on_SM_lp}) are identical to those in Table \ref{tab:generalized_pol}
obtained using the recurrence relations.

\section{Example 1: multisingular Gaussian beam\label{sec:Example-1}}

Now, let us exemplify the procedure of how to use the scattering modes
to solve the Schrödinger equation~(\ref{eq:par_diff_eq}) with a general 
initial condition showing a combination of singularities embedded
in a Gaussian beam. That is, we assume that the initial condition
is of the form 
\begin{equation}
\label{eq:initial_cond_a}
\phi(w,\overline{w},0)=\prod_{i=1}^{N_{a}}(w-a_{i})\prod_{i=1}^{N_{b}}(\overline{w}-b_{i})\phi_{00}(w,\overline{w},0),
\end{equation}
where $a_{i}$ ($b_{i}$) is the location of the $N_{a}$($N_{b}$)
positive (negative) singularities at $z=0$. We choose as an example
the case with two positive singularities out of the origin of the
transverse $(x,y)$ plane ---$a_{1},a_{2}\ne(0,0)$--- and one negative
singularity at the origin

\begin{eqnarray*}
\phi(w,\overline{w},0) & = & (w-a_{1})(w-a_{2})\overline{w}\phi_{00}(w,\overline{w},0).\\
\end{eqnarray*}
This is expanded as 
\begin{eqnarray*}
\phi(w,\overline{w},0) & = & \left[w|w|^{2}-(a_{1}+a_{2})|w|^{2}+a_{1}a_{2}\overline{w}\right]\phi_{00}.\\
\end{eqnarray*}
According to Eqs.~(\ref{eq:arbitrary_field-at-zero}) and (\ref{eq:expansion_t_coef_SM}),
and expressing $(n,\overline{n})$ as the quantum numbers $(l,p)$ by means
of the relations (\ref{eq:def_l}) and (\ref{eq:def_p}) one can obtain
the solution of Eq.~(\ref{eq:par_diff_eq}) valid for all $z$ in
terms of the scattering modes as

\begin{eqnarray}
\phi(w,\overline{w},z) & = & \Phi_{11}(w,\overline{w},z)-(a_{1}+a_{2})\Phi_{01}(w,\overline{w},z)+a_{1}a_{2}\Phi_{-10}(w,\overline{w},z).\label{eq:solution:ex2}\\
\nonumber 
\end{eqnarray}
Note the following simple rules in the procedure used to obtain solution
(\ref{eq:solution:ex2}): 
\begin{enumerate}
\item The coefficients of the scattering modes in the solution valid for all $z$ are \emph{identical} to those in the expansion in powers of $w$ and 
$\overline{w}$ in the initial condition.
\item Every power of $w$ or $\overline{w}$ gives the value of the angular
momentum quantum number $l$ in the corresponding scattering mode. 
\item Every power of $|w|^{2}$ gives the value of the radial quantum number
$p$. 
\end{enumerate}
The scattering modes $\Phi_{lp}$ are obtained from Eqs.~(\ref{eq:SM_complex_l_pos})
and (\ref{eq:SM_complex_l_neg}) upon substitution of the corresponding
values of $(l,p)$. The ones needed to evaluate Eq.~(\ref{eq:solution:ex2})
are
\begin{eqnarray}
\hspace{-1.7cm} \Phi_{11}\left(w,\overline{w},z\right) & =-w\left(\frac{z_{R}}{q(z)}\right)^{2}\left(\frac{2zz_{R}}{k_{0}q(z)}\right)\left(2-\gamma(z)|w|^{2}\right)\exp\left[-i\frac{k_{0}}{2}\frac{|w|^{2}}{q(z)}\right],\label{eq:SM_analytical-11}
\end{eqnarray}
\begin{eqnarray}
\hspace{-1.7cm}\Phi_{01}\left(w,\overline{w},z\right) & =\left(\frac{iz_{R}}{q(z)}\right)\left(\frac{2zz_{R}}{k_{0}q(z)}\right)\left(1-\gamma(z)|w|^{2}\right)\exp\left[-i\frac{k_{0}}{2}\frac{|w|^{2}}{q(z)}\right],\label{eq:SM_analytical-01}
\end{eqnarray}
\begin{eqnarray}
\hspace{-1.7cm}\Phi_{-10}\left(w,\overline{w},z\right) & =-\overline{w}\left(\frac{z_{R}}{q(z)}\right)^{2}\exp\left[-i\frac{k_{0}}{2}\frac{|w|^{2}}{q(z)}\right],\label{eq:SM_analytical-10}
\end{eqnarray}
where $q(z)\equiv z+iz_{R}$, $\gamma(z)=\left(k_{0}/2\right)z_{R}\left[zq(z)\right]^{-1}$
and 
we have restored the wavenumber $k_0$.
For each value of $z$ one can find analytically
the zeros of Eq.~(\ref{eq:solution:ex2}), after substituting (\ref{eq:SM_analytical-11})-(\ref{eq:SM_analytical-10}).
That is, one has to solve

\[
w\left(\frac{z_{R}}{q(z)}\right)\left[2-\gamma(z)|w|^{2}\right]+i(a_{1}+a_{2})\left[1-\gamma(z)|w|^{2}\right]+a_{1}a_{2}\overline{w}\left(\frac{k_{0}}{2z}\right)=0.
\]
This equation completely determines the trajectories of the singularities
for all values of $z$. Let us get a deeper insight by analyzing first
a symmetric case, that is, for example, when $a_{1}=-a_{2}$ , we
can get, from the previous expression 
\begin{equation}
wz_{R}\left[-4z(z+iz_{R})+k_{0} z_{R}w\overline{w}\right]+a_{1}^{2}k_{0} \overline{w}(z+iz_{R})^{2}=0,\label{eq:pos_sing_sym1}
\end{equation}
where we substituted the expressions for $q(z)$ and $\gamma(z)$
and took only its numerator. This has to be solved together with its
complex conjugate (we assume $a_{1}$ on the $x$ axis) 
\begin{equation}
\overline{w}z_{R}\left[-4z(z-iz_{R})+k_{0} z_{R}w\overline{w}\right]+a_{1}^{2}k_{0} w(z-iz_{R})^{2}=0.\label{eq:pos_sing_sym2}
\end{equation}
It is easy to check that, corresponding to the initial condition,
there is a root of these equations at $z=0$ at the origin with charge
$-1$ and two positive roots at $w=\pm a_{1}$. We can combine the
previous equations and find the $z$ at which the two positive roots
merge at the origin, obtaining that this occurs for 

\[
z_{\mbox{m}}=\pm\frac{a_{1}^{2}k_{0}z_{R}}{\sqrt{16z_{R}^{2}-a_{1}^{4}k_{0}^{2}}}.
\]

\begin{figure}
\begin{tabular}{cc}
(a) & (b)\tabularnewline
\includegraphics[width=7cm]{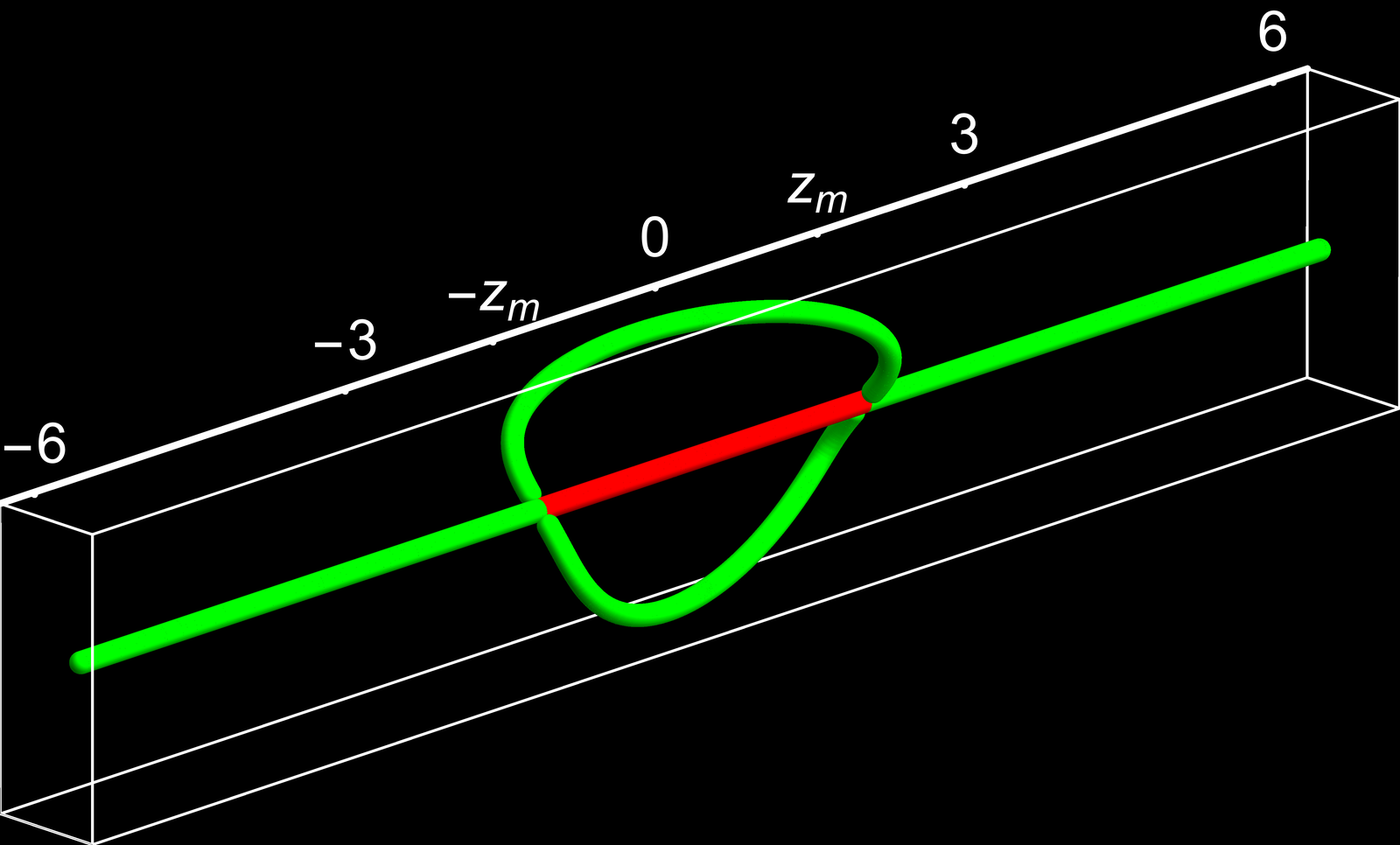} & \includegraphics[width=7.1cm]{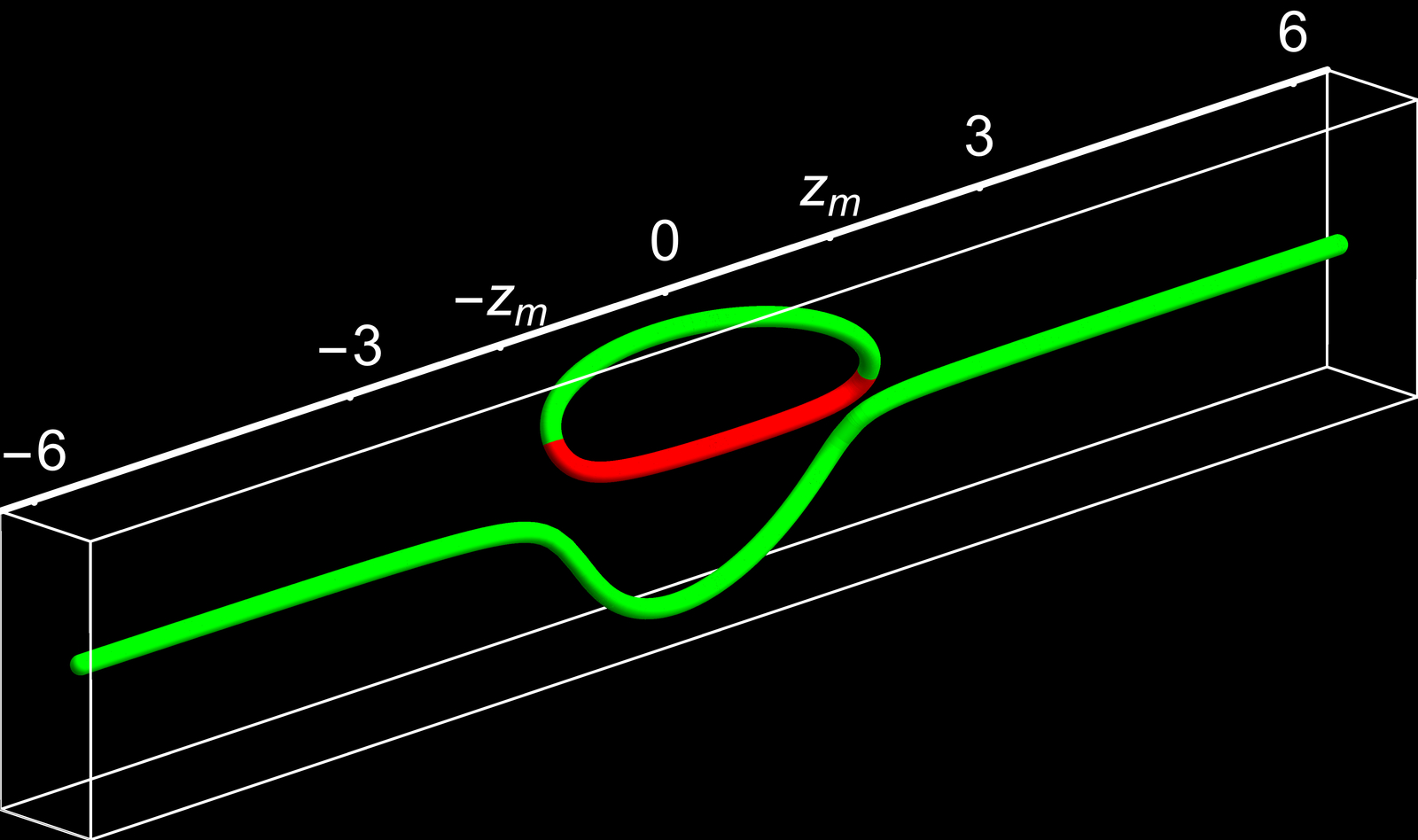}\tabularnewline
(c) & (d)\tabularnewline
\includegraphics[width=7cm]{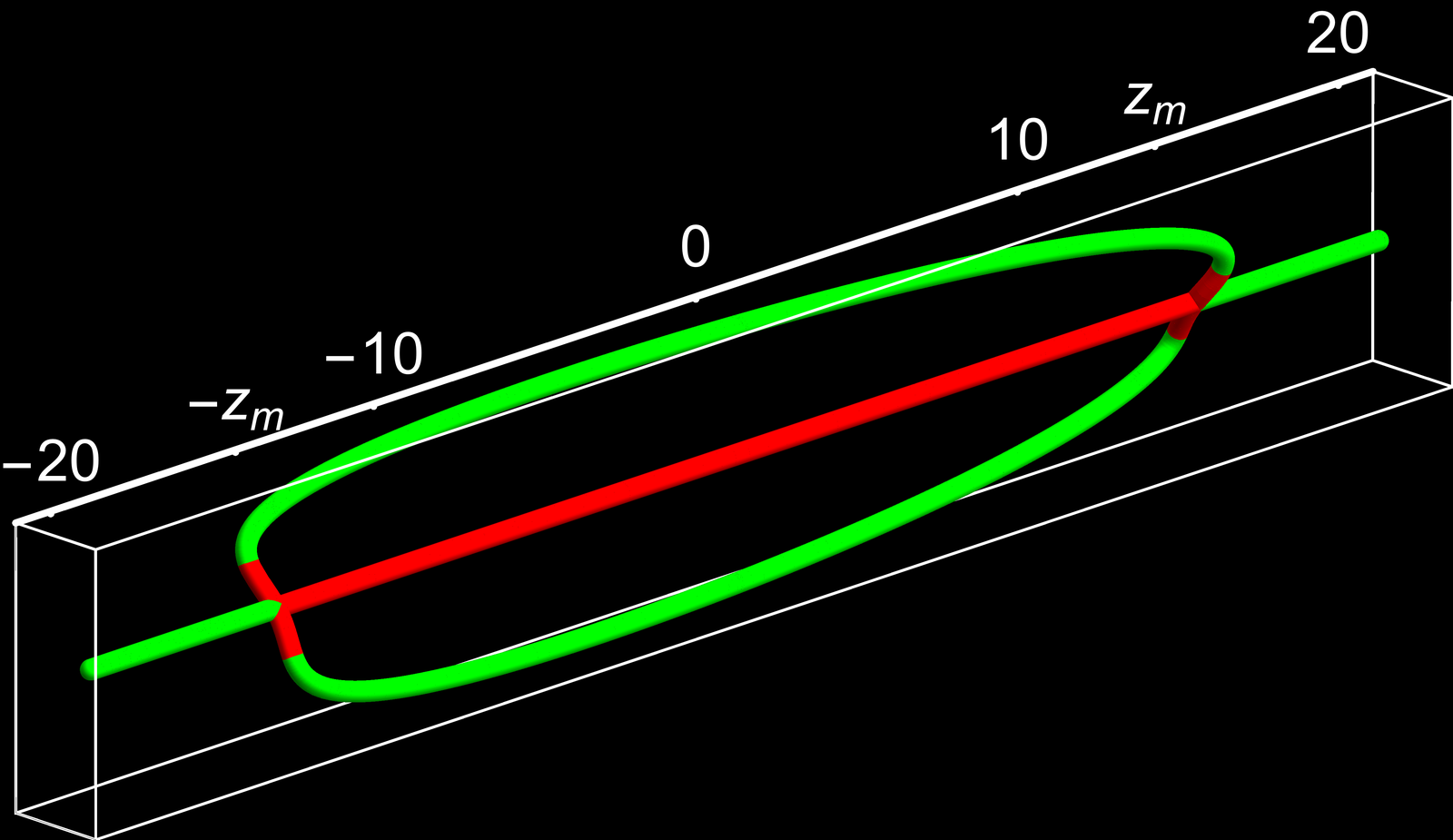} & \includegraphics[width=7.1cm]{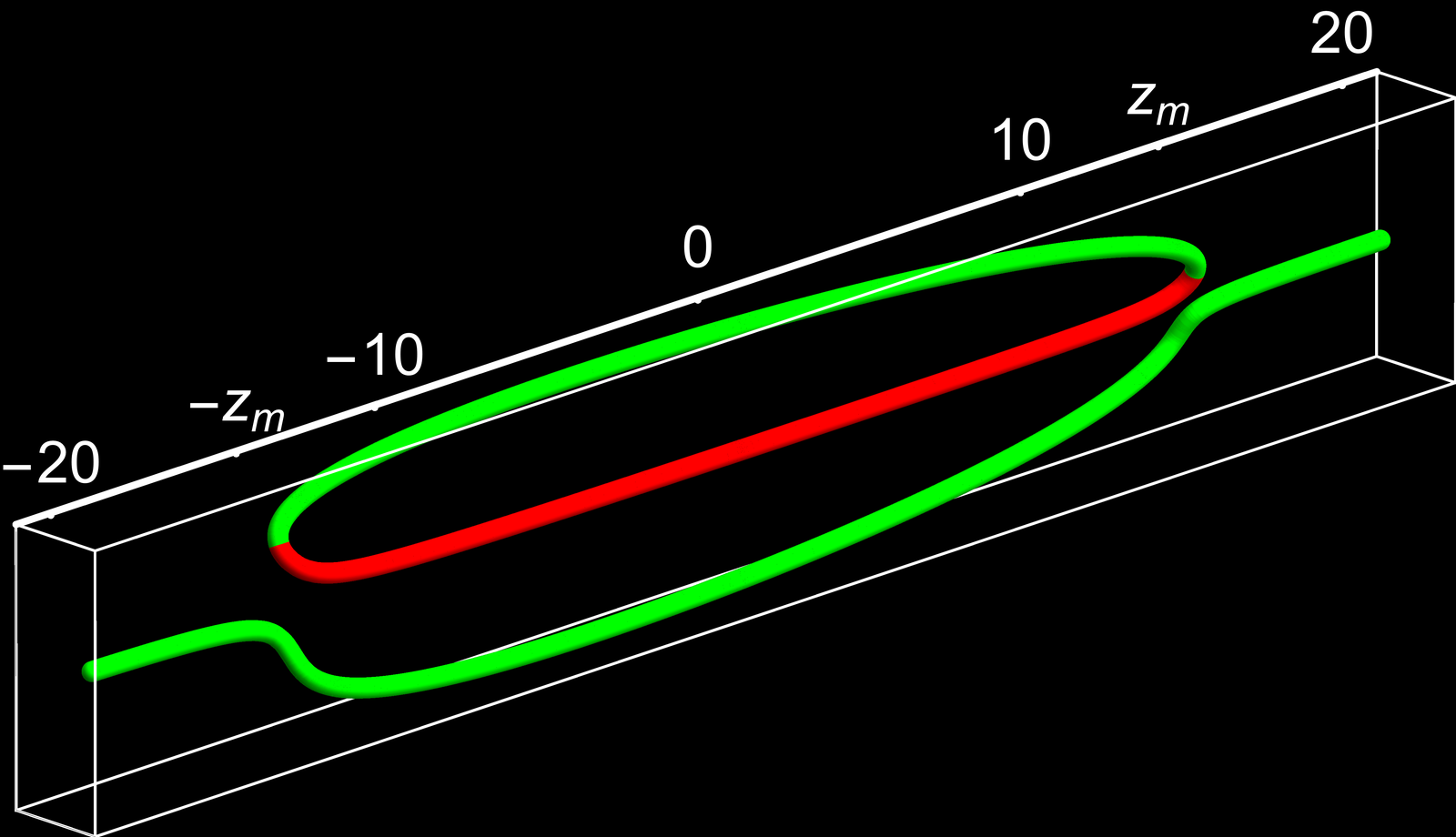}\tabularnewline
(e) & (f)\tabularnewline
\includegraphics[width=7cm]{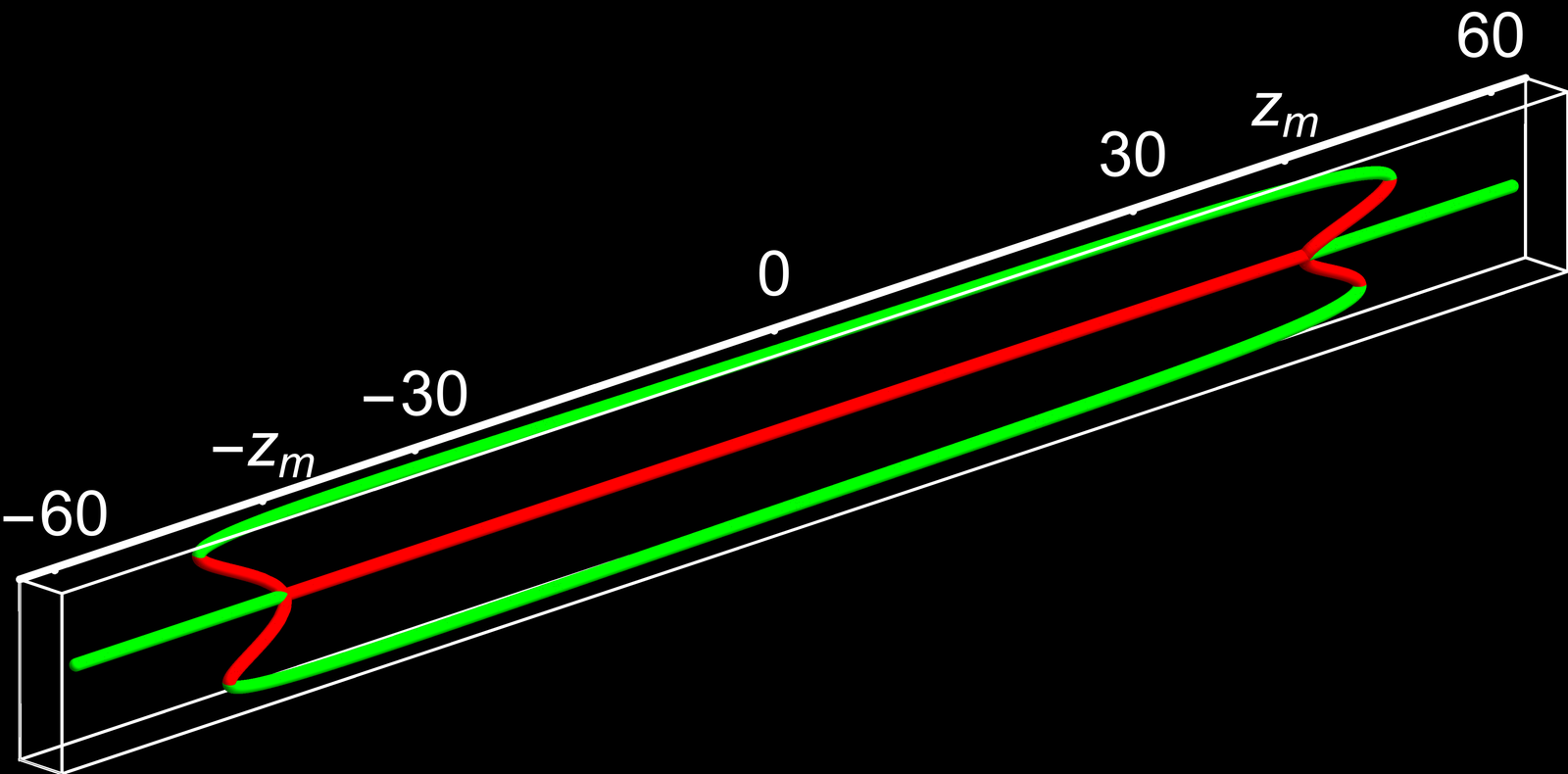} & \includegraphics[width=7.1cm]{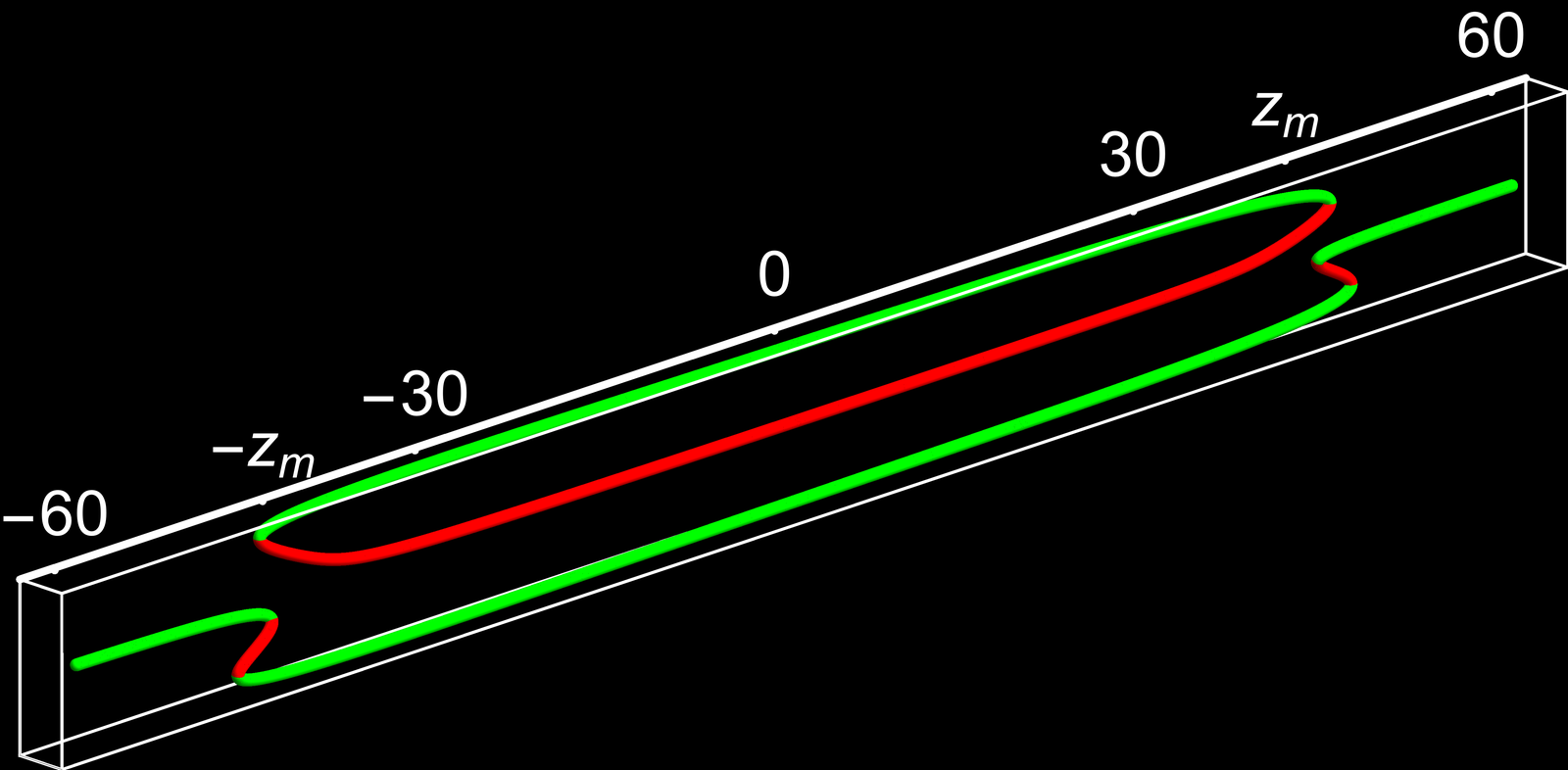}\tabularnewline
\end{tabular}
\caption{Vortex lines corresponding to a multisingular vortex Gaussian beam
at $z=0$ with two positively charged singularities at positions $a_{1}$
and $a_{2}$ different from the origin of the transversal $(x,y)$
plane, and one negatively charged at the origin. Green (red) lines
correspond to the position of the positive (negative) singularity
at each $z$. In the left column we represent the symmetric case,
that is, $a_{1}=-a_{2}$. Here, $a_{1}=1,3,5$ for (a), (c) , and
(e), respectively. In the right column we represent the slightly asymmetric
case, $a_{1}+a_{2}=\epsilon\protect\ne0$. The values for $a_{1}$
used in the symmetric case are also used in the right column, but
here $\epsilon=0.1$ (except for (f), were we use $\epsilon=0.2$
for a better pictorial representation). In all cases $z_{R}=100$
and $k_{0}=2\pi$. We indicate in each  panel the value of the merging
distance $z_{\mbox{m}}$. \label{fig:Vortex-lines}}
\end{figure}

Indeed, as shown in Fig. \ref{fig:Vortex-lines} (a), when the
positive singularities are located close to the origin ($a_{1}=1$),
the two positive charges located out of the origin merge with the
central, negative one, after a short evolution. Beyond this value
of $z_{\mbox{m}}$ there is only a single positive charge in the origin
(the same is valid for negative $z$). This merging occurs at $z_{\mbox{m}}=\pm1.571\approx\pm\pi/2$,
for the parameters used in the figure ($a_{1}=1$, $z_{R}=100$).
Note that when $z_{R}\gg a_{1}$ the expression for $z_{\mbox{m}}$
simplifies to approximately $\pm a_{1}^{2}\left(k_{0}/4\right)$,
which equals $\pm a_{1}^{2}\left(\pi/2\right)$ when using $k_{0}=2\pi$
normalization, elegantly explaining this behavior. Similarly occurs
when the initial condition fixes the off-axis singularities at a larger
distance from the origin [see Fig. \ref{fig:Vortex-lines} (c) and
(e)]. Note the non-trivial trajectories followed by the off-axis
singularities in the interval between the negative and positive values
of $z_{\mbox{m}}$. For the cases represented in Fig. \ref{fig:Vortex-lines}
(c) and (e) one observes that, close to the merging point, the central
negative charge produces two negative ones which move outwards and
cancel with the two positive off-axis ones, while leaving a central
one with changed sign. We also note that if $a_{1}^{4}k_{0}^{2}>16z_{R}^{2}$
no merging occurs, as we get an imaginary denominator. 

Let us now analyze a slightly asymmetric case, that is $a_{1}+a_{2}=\epsilon$.
Then we get that

\[
w\left(\frac{2zz_{R}}{k_{0}q(z)}\right)\left[2-\gamma(z)|w|^{2}\right]+i\epsilon\left(\frac{2z}{k_{0}}\right)\left[1-\gamma(z)|w|^{2}\right]+a_{1}(\epsilon-a_{1})\overline{w}=0
\]
together with its complex conjugate determine the position of the
singularities for all values of $z$. We have included in the right
column of Fig. \ref{fig:Vortex-lines} the same cases as in the
left column, but introducing a small asymmetry. The positive charge
located initially closer to the negative central one merges with it
at a distance approximately equal to $z_{\mbox{m}}$. The other
positive one bends and occupies a position close to the origin of
the $(x,y)$ plane after the merging point. In all cases the total
charge is conserved for all values of $z$ and equals $+1$.

\section{Example 2: discrete-Gauss state \label{sec:Example-2}}

A discrete-Gauss (DG) state is a solution of the paraxial diffraction
equation~(\ref{eq:par_diff_eq}), which, in polar coordinates, verifies
the initial condition
\begin{equation}
\phi_{lqNv}^{DG}(r,\theta,0)=\exp iv\left[2r^{N}\cos\left(N\theta\right)\right]\phi_{lq}^{LG}(r,\theta,0),\label{eq:initial_cond}
\end{equation}
$\phi_{lq}^{LG}$ being the mathematical expression of the Laguerre-Gauss
state $lq$. 

Physically, the exponential factor in Eq.~(\ref{eq:initial_cond})
corresponds to the amplitude transmittance function of a Diffractive
Optical Element (DOE) located at $z=0$. The form of its transmittance
function in complex coordinates ($w=x+iy=re^{i\theta}$, $\overline{w}=x-iy=re^{-i\theta}$)
\[
t=\exp iv\left[2r^{N}\cos\left(N\theta\right)\right]=\exp iv\left[w^{N}+\overline{w}^{N}\right]
\]
is the one of a DOE owning \emph{discrete} rotational symmetry of
order $N$. It represents the most general form of $t$ for a pure\emph{
discrete-symmetry} DOE (DSDE) close to the origin in the absence of
lensing effects (no dependence on $r^{2}=w\overline{w}$.) 

Explicit analytical expressions can be found for DG states at first
order in the $v$ parameter --known as the deformation parameter \cite{2014:Ferrando}.
At this order, any DG state can be written as a superposition of three
scattering modes [we consider here $q=0$ for the LG state in Eq.~(\ref{eq:initial_cond})]:
\begin{equation}
\phi_{_{lNv}}^{DG}\sim\begin{array}{cc}
\Phi_{l0}+iv\Phi_{l+N,0}+iv\Phi_{l-N,\overline{N}} & l\ge0\\
\Phi_{l0}+iv\Phi_{l+N,\overline{N}}+iv\Phi_{l-N,0} & l<0,
\end{array}\label{eq:DGS_superpos_3SM}
\end{equation}
where $\overline{N}=\min(|l|,N)$. The scattering modes   $\Phi_{lp}$
have the analytical expression (\ref{eq:SM_analytical}).

In this section we present an explicit example of the construction
of a DG state as a linear combination of scattering modes. This can be considered
a particular case of the general formula (\ref{eq:DGS_superpos_3SM}).
Nevertheless, in order to appreciate the simplicity of the construction
in terms of the scattering modes, it is instructive to provide an explicit example
solved from the very beginning. We consider the case of the DG state
$\phi_{lNv}^{DG}$ with $l=+3$ and $N=4$. The election of this state
is not coincidental since this was the case analytically solved in
Ref. \cite{2013:Ferrando} using a complete different method based
on the Fresnel diffraction integral. Moreover, this case was also
the one experimentally demonstrated in Ref. \cite{2014:Novoa} showing
an excellent agreement with the theoretical result.

From an experimental perspective, the $\phi_{+3,4,v}^{DG}$ state
is obtained from the LG mode $\phi_{+3,0}^{LG}$ by acting at its
waist (plane $z=0$) with a pure discrete-symmetry diffractive element
(DSDE) owning discrete rotational symmetry of order $N=4$ \cite{2014:Novoa}.
From a mathematical perspective, this operation defines the condition
fulfilled by the optical field at $z=0$, which near to the origin
can be written in the form (\ref{eq:initial_cond}). In complex notation
and to leading order in the deformation parameter $v$, the field
at $z=0$ is given by:
\begin{eqnarray*}
\phi_{+3,4,v}^{DG}\left(w,\overline{w},0\right) & = & \left[1+iv\left(w^{4}+\overline{w}^{4}\right)\right]w^{3}\bar{\phi}_{00}(|w|^{2},0)\\
 & = & \left[w^{3}+ivw^{7}+ivw^{3}\overline{w}^{4}\right]\bar{\phi}_{00}(|w|^{2},0)\\
 & = & \Phi_{3\overline{0}}\left(w,\overline{w},0\right)+iv\Phi_{7\overline{0}}\left(w,\overline{w},0\right)+iv\Phi_{3\overline{4}}\left(w,\overline{w},0\right).
\end{eqnarray*}
We see how the simple observation of the polynomial in brackets \emph{directly}
provides both the ``quantum numbers'' $n$ and $\overline{n}$ of the
three scattering modes involved \emph{and} the three components of the linear
combination. In fact, the problem has been already automatically solved
since the \emph{same} linear combination provides also the solution
for the whole space:
\begin{eqnarray*}
\phi_{+3,4,v}^{DG}\left(w,\overline{w},z\right) & = & \Phi_{3\overline{0}}\left(w,\overline{w},z\right)+iv\Phi_{7\overline{0}}\left(w,\overline{w},z\right)+iv\Phi_{3\overline{4}}\left(w,\overline{w},z\right)\\
 & = & \Phi_{30}\left(w,\overline{w},z\right)+iv\Phi_{70}\left(w,\overline{w},z\right)+iv\Phi_{-1,3}\left(w,\overline{w},z\right),
\end{eqnarray*}
where in the last step we have changed the scattering modes representation in
terms of the $(n,\overline{n})$ ``quantum numbers'' into the one
in terms of the $(l,p)$ numbers, according to the relations (\ref{eq:def_l})
and (\ref{eq:def_p}). The later result provides an analytical representation
of the DG state using the explicit equations for the scattering modes in complex
coordinates (\ref{eq:SM_complex_l_pos}) and (\ref{eq:SM_complex_l_neg})
along with the expressions for $F$-polynomials given in previous
sections. 
\[
\hspace{-2.4cm}\phi_{+3,4,v}^{DG}\left(w,\overline{w},z\right)\!=\!\left\{ w^{3}\alpha^{4}(z)\!+\!ivw^{7}\alpha^{8}(z)\!+\!iv\overline{w}\alpha^{2}(z)\beta^{3}(z)F_{3}^{1}\left[\gamma(z)|w|^{2}\right]\right\}\! \exp\left[-i\frac{k_{0}}{2}\frac{|w|^{2}}{q(z)}\right]\!,
\]
where
\begin{eqnarray*}
\alpha(z) & = & \frac{iz_{R}}{q(z)},\\
\beta(z) & = & \frac{2zz_{R}}{k_{0}q(z)},\\
\gamma(z) & = & \frac{k_{0}z_{R}}{2zq(z)}
\end{eqnarray*}
and
\[
F_{3}^{1}(x)=24-36x+12x^{2}-x^{3}.
\]
In order to compare with the result obtained for the same case in
Ref. \cite{2013:Ferrando}, we rewrite the previous expression as:
\begin{eqnarray}
\hspace{-1.7cm}\phi_{+3,4,v}^{DG}\left(w,\overline{w},z\right) & = & \alpha{}^{8}\left[w^{3}\alpha^{-4}+ivw^{7}+iv\overline{w}\alpha{}^{-6}\beta^{3}F_{3}^{1}\left(\gamma|w|^{2}\right)\right]\exp\left[-i\frac{k_{0}}{2}\frac{|w|^{2}}{q(z)}\right]\nonumber \\
\hspace{-1.7cm} & = & \alpha^{8}\left[A_{0}(z)w^{3}+A_{+}w^{7}+A_{-}(|w|^{2},z)\overline{w}\right]\exp\left[-i\frac{k_{0}}{2}\frac{|w|^{2}}{q(z)}\right],\label{eq:final_expres_SM_3-4}
\end{eqnarray}
in which
\begin{eqnarray*}
A_{0}(z) & = & \alpha^{-4}=\frac{q(z)^{4}}{z_{R}^{4}},\\
A_{+} & = & iv
\end{eqnarray*}
and
\begin{eqnarray*}
A_{-}(|w|^{2},z) & = & iv\alpha^{-6}\beta^{3}F_{3}^{1}\left(\gamma|w|^{2}\right)\\
 & = & iv\alpha^{-6}\beta^{3}\gamma^{3}\left[24\gamma^{-3}-36\gamma^{-2}|w|^{2}+12\gamma^{-1}|w|^{4}-|w|^{6}\right]\\
 & = & iv\left[-24\overline{q}^{3}+36\overline{q}^{2}|w|^{2}-12\overline{q}|w|^{4}+|w|^{6}\right],
\end{eqnarray*}
where in the last step we have used the identity $\gamma=-\alpha^{2}/\beta$
and the definition $\overline{q}\equiv\gamma^{-1}$ used in \cite{2013:Ferrando}
for comparison purposes. Our final result (\ref{eq:final_expres_SM_3-4})
is identical to that obtained in the aforementioned reference using
a complete different method based on the non-trivial calculation of
the Fresnel diffraction integral.

\section{Conclusions\label{sec:Conclusions}}

In this article we have presented the mathematical theory for the
systematic and analytical calculation of the so-called scattering
modes previously defined in Ref. \cite{2014:Ferrando}. Additionally,
this mathematical construction permits to provide a simple and efficient
procedure to analytically determine discrete-Gauss states, also defined
and introduced in the same reference. We have introduced a systematic
construction of the so-called fundamental and generalized $F$-polynomials,
which determine completely the analytical form of any scattering mode
and, thus, indirectly, of any discrete-Gauss state or of any superposition
of scattering modes. Using the concepts of raising and lowering operators
we have found the recurrence relations that $F$-polynomials satisfy.
These recurrence relations, both directly for polynomials themselves
or for their coefficients, permit to access to analytical expressions
for the scattering modes in an analogous way as for standard  mode sets,
such as Laguerre-Gauss modes. In this sense, $F$-polynomials play
a similar role as Laguerre polynomials do for Laguerre-Gauss modes.
As already pointed in Ref. \cite{2014:Ferrando}, scattering modes
constitute a basis for the expansion of any paraxial optical free
field. However, they are not orthogonal nor bi-orthogonal as Laguerre-Gauss
or discrete-Gauss modes are. Despite this apparent drawback, scattering
modes are an excellent basis to expand free propagating optical fields
hosting phase singularities embedded in Gaussian envelopes. As shown
in the explicit examples provided in this article, the knowledge of
the singularity structure of a beam, basically its polynomial structure
in terms of the complex variables $w$ and $\overline{w}$ at a given
propagation plane, determines in a \emph{simple and direct} form the
components of the beam in the scattering mode basis. This linear combination
is the solution in the entire space\emph{ }and its construction does
not require any type of projection\emph{ }operation\emph{. }The determination
of the whole solution is thus extremely simple and it is obtained
in an analytical form with the help of the provided explicit construction
of the $F$-polynomials. In this direction, in Section \ref{sec:Example-1}
we provide a neat example of the analytical calculation of a multisingular
vortex Gaussian beam in terms of scattering modes starting exclusively
from the knowledge of its multisingular structure at a given plane.
This analytical resolution permits to obtain closed analytical expressions
for meaningful quantities such as the merging axial distances at which
vortex-anti vortex loops are created and annihilated. In addition,
in Section \ref{sec:Example-2} we present an example of calculation
of an specific discrete-Gauss state previously obtained using the
integral Fresnel diffraction method. Remarkably, here the same problem
is solved in a simple an elegant way without the need to resort to
any integration nor projection operation. 
The method presented here can help to bring a different perspective to relevant optical problems such as those related to random light fields, like in the phenomenon of laser speckle. In laser speckle fields, phase singularities lines form intricate tangles and/or knots. This context is promising to apply the present procedure since the initial condition of the speckle field is given by a random distribution of singularities, which can be easily modeled by allowing the positions of the singularities, given by the coefficients  $b_i$ and $a_i$ in Eq.~(\ref{eq:initial_cond_a}), to be distributed randomly. The  main drawback is that the resulting set of equations may not be tractable analytically. The advantage is that it translates the problem of solving the equation in the problem of finding the zeros of a high order polynomial, which can always be computed numerically.

We finally emphasize here
that, even though we presented this method in the context of optics,
it can be equally used in any field in which it is required to solve
the Schrödinger equation with an initial condition formally identical
to a multisingular vortex Gaussian beam. 
This is the case already mentioned of ultra-cold atoms forming a BEC in magnetic traps, described by an equation formally identical to the nonlinear Schrödinger equation. In this BEC system, the authors already discussed how an initially highly charged vortex is annihilated by the action of a discrete symmetry potential, exactly as in the system described in~\cite{2013:Ferrando}, with the main difference being here the presence of an additional parabolic trapping and a Kerr-defocusing nonlinearity ~\cite{2012CommefordPRA}. 
Another alternative  application of this method can be found in the framework of electron quantum mechanics. Very recent experiments with electron-vortex beams in interaction with apertures of different discrete symmetries unveil a very rich multisingular electron phase structure completley analogous to that found for light beams in optics ~\cite{2016ClarkarXiv}. The mathematical mapping between the Schrödinger equation and the paraxial wave equation permits to establish a direct connection between the present formalism and experimental results of this type.

Finally, as an outlook, while in its present form the method cannot be applied to random field statistical mechanics systems, such as in the random $x$-$y$ or the Ising model, where the concept of vortex line is also of fundamental importance~\cite{1975ImryPRL,1996PRBGingras,2013GaraninEPL}, we believe that future research following the lines discussed here will give valuable insight for this context.

\ack
This work was supported by the MINECO (Government of Spain) under
Grants No. TEC2010-15327, TEC2013-50416-EXP and TEC2014-53727-C2-1-R.
M.A. G-M. acknowledges support from EU grants OSYRIS (ERC-2013-AdG
Grant No 339106), SIQS (FP7-ICT-2011-9 No 600645), EU STREP QUIC (H2020-FETPROACT-
2014 No 641122), EQuaM (FP7/2007 \textendash{} 2013 Grant No 323714),
Spanish Ministry grant FOQUS (FIS2013-46768-P), the Generalitat de
Catalunya project 2014 SGR 874, and Fundació Cellex. Financial support
from the Spanish Ministry of Economy and Competitiveness, through
the \textquoteleft{}Severo Ochoa\textquoteright{} Programme for Centres
of Excellence in R\&D (SEV-2015-0522) is acknowledged.

\section*{References}
\providecommand{\newblock}{}

\end{document}